\documentclass[british]{article}
\usepackage{mathptmx}

\usepackage[T1]{fontenc}
\usepackage[latin9]{inputenc}
\usepackage{geometry}
\usepackage{units}
\usepackage{bm}
\usepackage{amsmath}
\usepackage{amssymb}
\usepackage{graphicx}
\usepackage{setspace}
\usepackage[authoryear]{natbib}
\makeatletter

\providecommand{\tabularnewline}{\\}

\newcommand{\lyxaddress}[1]{
\par {\raggedright #1
\vspace{1.4em}
\noindent\par}
}

\usepackage{bm}

\makeatother

\usepackage{babel}
\begin{document}

\title{Population genetics and substitution models of adaptive evolution}

\author{Mario dos Reis}

\maketitle

\lyxaddress{mario.barros@ucl.ac.uk\\
The MRC National Institute for Medical Research, Mill Hill, London,
UK, NW7 1AA\\
Current address: Department of Genetics, Evolution and Environment,
University College London, London, UK, WC1E 6BT}

Keywords: adaptive evolution, population genetics, substitution model,
selection, influenza, fixation probability

Running head: Population genetics of adaptive evolution

\lyxaddress{\pagebreak{}}

\section*{Abstract}

The ratio of non-synonymous to synonymous substitutions $\omega(=d_{N}/d_{S})$
has been widely used as a measure of adaptive evolution in protein
coding genes. Omega can be defined in terms of population genetics
parameters as the fixation ratio of selected vs. neutral mutants.
Here it is argued that approaches based on the infinite sites model
are not appropriate to define $\omega$ for single codon locations.
Simple models of amino acid substitution with reversible mutation
and selection are analysed, and used to define $\omega$ under several
evolutionary scenarios. In most practical cases $\omega<1$ when selection
is constant throughout time. However, it is shown that when the pattern
of selection on amino acids changes, for example after an environment
shift, a temporary burst of adaptive evolution ($\omega\gg1$) can
be observed. The fixation probability of a novel mutant under frequency
dependent selection is calculated, and it is used to show why $\omega>1$
can be sometimes expected for single locations at equilibrium. An
example with influenza data is discussed.\pagebreak{}

\section*{Introduction}

Imagine a population with $N$ individuals. In one particular generation
an individual with a novel, selectively neutral mutation is born.
Initially there is only one copy of the novel mutant allele in the
population. If the organism is diploid, the initial frequency of the
allele is $\nicefrac{1}{2N}$. Stochastic fluctuations (random drift)
will govern the fate of the mutant in the population. A classical
result from population genetics theory is that the probability (after
infinitely many generations) that \emph{all} individuals will eventually
carry the mutation is exactly $\nicefrac{1}{2N}$. In this case we
say the mutant allele has become fixed. If $\mu$ neutral mutations
appear per genome per generation, then in a single generation, $2N\mu$
mutant alleles would be produced, each with a $\nicefrac{1}{2N}$
chance of ultimately spreading throughout the population. Thus $2N\mu\nicefrac{1}{2N}=\mu$
is the rate at which neutral mutant alleles become fixed in the population
each generation. This is a classical result from the neutral theory
of molecular evolution, which maintains that most changes at the molecular
level in populations are due to the random fixation of selectively
neutral mutants (\citealt{Kimura1968,Kimura1983}).

Natural selection has an important effect on the probability of ultimate
fixation of a novel mutant (\citealt{Kimura1962}). If the mutant
is advantageous, its probability of fixation will be larger than that
of neutral mutants. Thus, regions of a genome generating advantageous
mutants will be substituted at a faster rate than equivalent regions
that evolve neutrally. On the other hand, if novel mutants are deleterious,
their probability of fixation will be lower than that of neutral mutants,
and regions of a genome that produce deleterious mutants will be substituted
at a slower rate than equivalent regions that evolve neutrally. This
provides the theoretical basis to detect adaptive evolution in protein
coding genes. If we assume that synonymous sites evolve neutrally
and that non-synonymous sites are under the influence of natural selection,
then the ratio of non-synonymous to synonymous substitutions $\omega(=d_{N}/d_{S})$
will reflect the dominant type of selection acting on the protein.
Under this interpretation, $\omega>1$, $\omega=1$ and $\omega<1$
indicate proteins under positive (diversifying) selection, neutral
evolution and negative (purifying) selection respectively. 

Very sophisticated methods have been developed to estimate $\omega$
from sequence alignments, methods that take into account the fact
that synonymous and non-synonymous sites are present in unequal numbers
and that the mutation rates for both types of sites are different
(synonymous mutations are usually transitions, which occur at a faster
rate than transversions, the dominant type of non-synonymous mutation).
The literature describing statistical techniques to estimate $\omega$
is extensive (e.g.~\citealt{Miyata+Yasunaga1980,Nei+Gojobori1986,Goldman+Yang1994,Ina1995};
for reviews see \citealt{Yang+Bielawski2000}; and ch~8 in \citealt{Yang2006})
and a large number of works have been published detecting positive
selection in, for example, viral coating proteins (\citealt{Fitch+1997,Yang2000,Yang+2003}),
in the human histocompatibility complex (\citealt{Hughes+Nei1988,Yang+Swanson2002}),
primate lysozymes (\citealt{Messier+Stewart1997,Yang1998}), etc (\citealt{Endo+1996}).

For most protein coding genes, only a few locations are expected to
be under positive selection, with the rest of the locations expected
to be under strong purifying constraints. Hence, statistical methods
that aim to detect positive selection and that consider a single $\omega$
value for a protein are expected to be conservative. Considerable
work has thus been aimed at models that consider variable $\omega$
ratios among sites (\citealt{Nielsen+Yang1998,Yang+2000,Huelsenbeck+2006}),
and tests have been developed to detect single codon locations were
$\omega>1$. Here there is a subtle but important point about the
biological meaning of a codon site where $\omega>1$. The theoretical
justification discussed above actually assumes that genomes are composed
of an infinite number of sites, and mutations do not occur more than
once at any given location. The substitution rate of a genomic region
under positive selection can only be larger than the substitution
rate of a neutral region if advantageous mutants are persistently
generated. The problem is that, when the number of sites is finite,
the substitution rate per site would eventually decay to the level
seen at sites under purifying selection. This is because once an advantageous
mutant has become fixed, any new mutations at the location would then
be deleterious. Under a finite sites model with reversible mutation
further assumptions about the nature of selection are needed to understand
$\omega>1$.

The methods developed so far do not seem to have attempted to construct
$\omega$ in terms of population genetic parameters. In my opinion,
there is some weakness between the theoretical interpretation of $\omega$
(such as in the discussions by \citealt{Bustamante2005}, or \citealt{Nielsen+Yang2003})
and the development of codon substitution models to detect positive
selection. In this work I explore the population genetics interpretation
of $\omega$ under simple models of amino acid substitution with reversible
mutation and selection. Several evolutionary scenarios, such as time
non-homogeneous selection and frequency dependent selection are discussed.
Some simple examples with viral data are presented, and the contribution
of different types of selection to the evolution of flu viruses is
discussed.

\section*{Theory}

In the following discussion we assume a Fisher-Wright model of random
genetic drift (e.g.~\citealt{Wright1931}). We work with idealised
populations where the effective and the real population numbers are
the same. Locations in a gene (or genome) are assumed to evolve independently,
and they do not interfere with each other. We assume genic selection
and that the selection coefficients involved in the models are small,
so that some simplifying approximations about fixation probabilities
can be made (\citealt{Kimura1962}). It is also assumed that mutation
rates are sufficiently small so that polymorphism is negligible and
locations remain fixed most of the time. This is particularly important
for the reversible mutation models described below (\citealt{Bulmer1991}).
The evolutionary process is viewed over long periods so the time from
appearance to fixation of a novel mutant is nearly instantaneous.
Issues such as selection on codon usage or non-homogeneous mutation
patterns are ignored. These assumptions are necessary to simplify
the mathematical treatment of the models discussed below.

\subsection*{Infinite sites model}

We can use Kimura's infinite sites model (\citealt{Kimura1969}; p.~46
in \citealt{Kimura1983}) to find an analytical expression for $\omega$.
Let's consider a large genome where the mutation rate per haploid
genome per generation is $\mu$. Because the genome is very large,
we assume that all new mutants always appear at new locations, effectively
assuming that there is no reversible mutation. Of all the new mutants
produced every generation a fraction $f_{0}\mu$ are neutral, and
a fraction $f_{s}\mu$ are selected with selection coefficient $s$
($f_{0}+f_{s}=1$). There are $N$ haploid genomes in the population
and the probability of ultimate fixation of a newly arisen mutant
is given by $u(S)=S/N(1-e^{-S})$, where $S=2Ns$ is the confounded
selection coefficient (\citealt{Fisher1930,Wright1931,Kimura1962}).
We note that the fixation probability of a neutral mutant is $u(0)\equiv\lim_{S\rightarrow0}u(S)=1/N$.
Thus each generation, $k_{s}=Nf_{s}\mu u(S)$ selected, and $k_{0}=Nf_{0}\mu/N$
neutral mutants are produced that will become ultimately fixed. Thus
the expected substitution (fixation) rate per generation is given
by
\[
k=k_{s}+k_{0}=Nf_{0}\mu u(0)+Nf_{s}\mu u(S).
\]

We are interested in defining $\omega$ in terms of the relative fixation
rates of selected vs. neutral mutants, and this is 
\begin{equation}
\omega=\frac{k_{s}}{k_{0}}\times\frac{f_{0}}{f_{s}}=\frac{\mu f_{s}u(S)}{\mu f_{0}u(0)}\times\frac{f_{0}}{f_{s}}=\frac{S}{1-e^{-S}},\label{eq:simple-omega}
\end{equation}
where the $f_{0}/f_{s}$ term is included to normalise $\omega$ (\citealt{Nielsen+Yang2003,Bustamante2005}).
The right hand side of equation~\ref{eq:simple-omega} is simply
the relative fixation probability of selected vs. neutral mutants:
$h(S)\equiv u(S)/u(0)=S/(1-e^{-S})$. Thus $\omega>1$, $=1$, and
$<1$ represent positive ($S>0$), neutral ($S=0$) and negative ($S<0$)
selection. Note that for large $S$ we have that $h(S)\approx S$. 

Equation~\ref{eq:simple-omega} defines $\omega$ in terms of the
fraction of neutral and selected \emph{mutants} that appear in a genome
each generation. We can also define $\omega$ in terms of the numbers
of neutral and selected \emph{sites} present. If there are fractions
$f_{s}^{*}$ of selected and $f_{0}^{*}$ neutral sites, each producing
mutants at rates $\mu_{s}$ and $\mu_{0}$ respectively, then the
total number of mutant alleles produced per generation is $\mu\equiv f_{s}^{*}\mu_{s}+f_{0}^{*}\mu_{0}$.
The fraction of selected mutants is then $f_{s}=f_{s}^{*}\mu_{s}/(f_{s}^{*}\mu_{s}+f_{0}^{*}\mu_{0})$
and the fraction of neutral mutants is $f_{0}=1-f_{s}^{*}$. It follows
that $\mu f_{s}=\mu_{s}f_{s}^{*}$ and $\mu f_{0}=\mu_{0}f_{0}^{*}$.
Thus, an equivalent definition of $\omega$ is
\[
\omega=\frac{\mu_{s}f_{s}^{*}u(S)}{\mu_{0}f_{0}^{*}u(0)}\times\frac{\mu_{0}f_{0}^{*}}{\mu_{s}f_{s}^{*}}=h(S),
\]
where the $\mu_{0}f_{0}^{*}/\mu_{s}f_{s}^{*}$ normalising term takes
into account the fact that there are different numbers of neutral
and selected sites, each producing mutants at different rates.

We can imagine a protein coding gene as being composed of a fraction
$f_{0}^{*}$ of synonymous, neutrally evolving sites and a fraction
$f_{s}^{*}$ of non-synonymous sites constrained by natural selection.
If the protein is composed of thousands of sites, and if we observe
the evolution of this protein during a short period of time, so that
backward mutations are unlikely, then equation~\ref{eq:simple-omega}
summarises the evolutionary dynamics and the selective pressure acting
on the protein.

As noted in the introduction, this interpretation of $\omega$ as
a measure of positive selection is conservative, especially since
in real proteins we might have fractions $f_{s_{-}}$ and $f_{s_{+}}$of
negatively and positively selected mutants arising each generation
($s_{-}<0$, $s_{+}>0$; $f_{s_{-}}+f_{s_{+}}+f_{0}=1$). Defining
$\omega$ in terms of $f_{s_{-}}$and $f_{s_{+}}$ we have that
\begin{equation}
\omega=\frac{k_{s_{-}}+k_{s_{+}}}{k_{0}}\times\frac{f_{0}}{f_{s_{-}}+f_{s_{+}}}=\frac{f_{s_{-}}h(S_{-})+f_{s_{+}}h(S_{+})}{f_{s_{-}}+f_{s_{+}}},\label{eq:average-omega}
\end{equation}
which is simply the weighted average of the relative fixation probabilities
of both types of mutants. Thus, if there is a large fraction of negatively
selected mutants, we might expect $\omega<1$ even if $S_{+}\gg1$
(formally $\omega<1$ if $f_{s_{-}}(1-h(S_{-}))>f_{s_{+}}(h(S_{+})-1)$).
Equation~\ref{eq:average-omega} can be easily extended to any arbitrary
number of classes of selected mutants.

Given the conservative nature of equation~\ref{eq:average-omega},
it would be better if we could explicitly define $\omega$ for each
codon location in a gene. The infinite sites model specifically ignores
reversible mutation. Under this model a new mutant codon is either
fixed or lost, and if it becomes fixed, it will remain in that state,
rendering the site specific substitution rate effectively equal to
zero. Equation~\ref{eq:simple-omega} is a useful approximation for
interpreting the relative evolutionary rates of different proteins,
or even to compare the rates of the same protein along different lineages
in a phylogenetic tree, however, it is not appropriate to understand
$\omega$ for single locations. An explicit finite sites model with
reversible mutation is needed. The reader can consult \citet{Bustamante2005}
for a discussion of $\omega$ under the infinite sites model.

\subsection*{A simple amino acid substitution model with selection and reversible
mutation}

We now focus our attention to defining $\omega$ for a single amino
acid location in a protein. The tertiary structure of a protein imposes
constraints on which amino acids can be accommodated at a particular
location. We focus our interest in locations where only two amino
acids, $c_{1}$ and $c_{2}$, are observed. Any mutant different from
$c_{1}$ and $c_{2}$ is lethal and hence can never become fixed in
the population. The neutral mutation rate per genome per generation
from $c_{1}$ to $c_{2}$ is $\alpha$, and $\beta$ is the rate in
the opposite direction ($\alpha,\beta>0$). The selection coefficient
in favour of $c_{1}$ is $s(>0)$ and against $c_{2}$ is $-s$. When
the population is fixed for $c_{1}$, the substitution rate ($q_{12}$)
from $c_{1}$ to $c_{2}$ is $N\alpha u(-S)=\alpha h(-S)$. Similarly
$\beta h(S)$ is the rate ($q_{21}$) from $c_{2}$ to $c_{1}$ when
the population is initially fixed for $c_{2}$. The substitution rates
($q_{ij}$) can be accommodated into the rate matrix
\begin{equation}
\mathrm{\mathbf{Q}}=\left(\begin{array}{cc}
-\alpha h(-S) & \alpha h(-S)\\
\beta h(S) & -\beta h(S)
\end{array}\right),\label{eq:Q-simple2x2}
\end{equation}
where the diagonal elements satisfy $q_{ii}=-\sum_{j\neq i}q_{ij}$.
The probability that the location, currently fixed for $c_{i}$, will
become fixed for $c_{j}$ after a certain time $t$ (in generations)
has elapsed is given by the transition probability matrix $\mathrm{\mathbf{P}}_{t}=\{p_{ij}(t)\}=e^{t\mathrm{\mathbf{Q}}}$.
Equation~\ref{eq:Q-simple2x2} describes the pattern of amino acid
substitution at the location as a time continuous Markov process.

During its evolutionary history, the location will spend a fraction
of time $\pi_{1}$ fixed at $c_{1}$ and a fraction $\pi_{2}$ at
$c_{2}$ ($\pi_{1}+\pi_{2}=1).$ At equilibrium the balance equation
\begin{equation}
\pi_{1}\alpha h(-S)=\pi_{2}\beta h(S)\label{eq:balance-eq}
\end{equation}
 holds. Solving for $\pi_{1}$ and $\pi_{2}$ we have $\pi_{1}=\beta h(S)/(\beta h(S)+\alpha h(-S))$
and $\pi_{2}=\alpha h(-S)/(\beta h(S)+\alpha h(-S)).$ Parameters
$\pi_{1}$ and $\pi_{2}$ can also be interpreted as the expected
equilibrium frequencies of amino acids $c_{1}$ and $c_{2}$ for a
collection of locations under the same substitution (selection + mutation)
pattern. 

Setting $\nu=\beta h(S)+\alpha h(-S)$ is easy to show that $\mathrm{\mathbf{Q}}$
can be written as

\begin{equation}
\mathrm{\mathbf{Q}}=\left(\begin{array}{cc}
-\nu\pi_{2} & \nu\pi_{2}\\
\nu\pi_{1} & -\nu\pi_{1}
\end{array}\right),\label{eq:Q-nu-pi}
\end{equation}
which is the simplest $2\times2$ reversible amino acid substitution
matrix. Parameter $\nu$ would then represent the `exchangeability'
of $c_{1}$ and $c_{2}$.

The expected substitution rate at equilibrium for the location is
given by

\begin{eqnarray*}
k_{s} & = & \pi_{1}\alpha h(-S)+\pi_{2}\beta h(S)\\
 & = & \frac{2\alpha\beta h(S)h(-S)}{\beta h(S)+\alpha h(-S)}.
\end{eqnarray*}

\begin{figure}[p]
\begin{centering}
\includegraphics[scale=0.8]{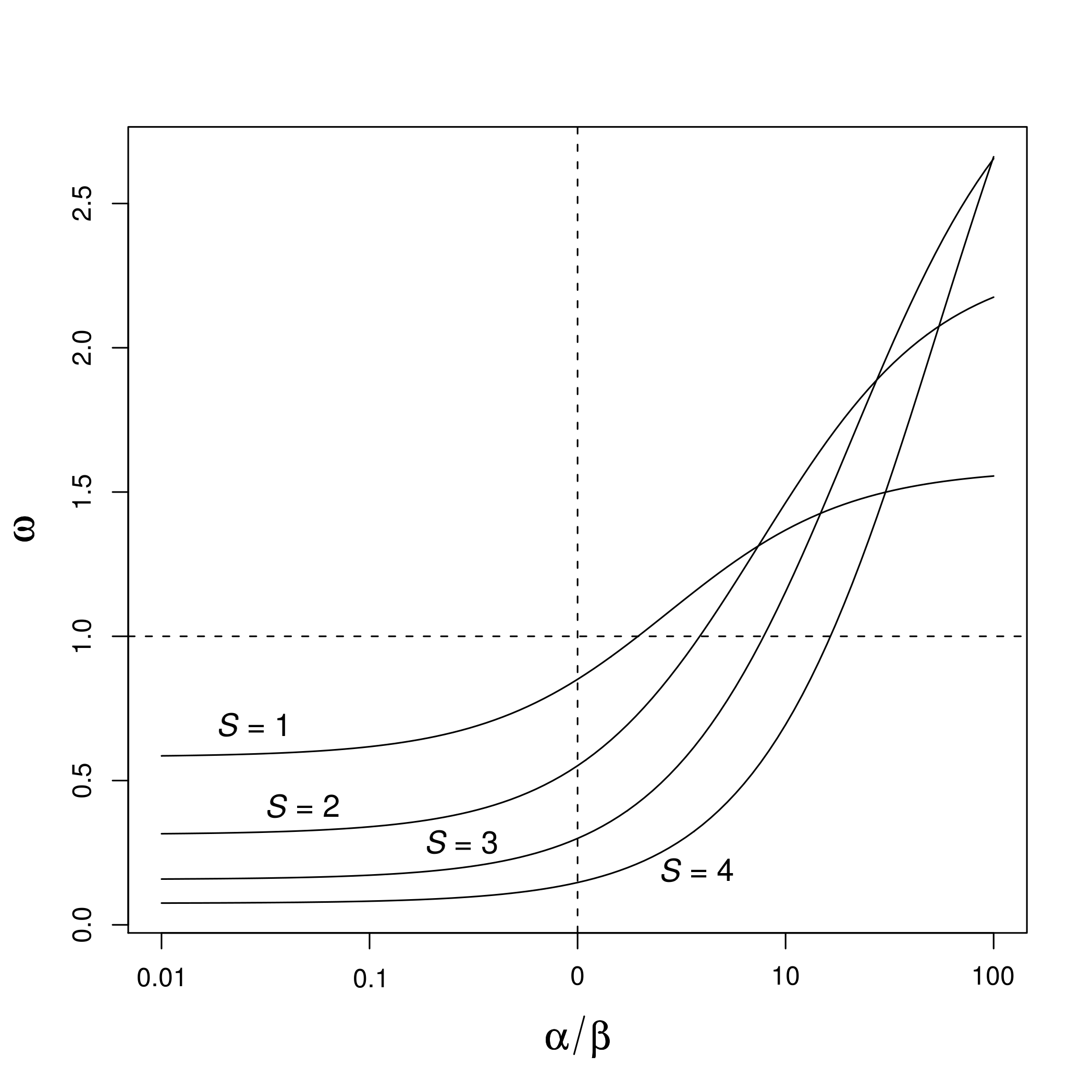}
\par\end{centering}

\caption{\label{fig:ab-omega}The expected value of $\omega$ in a simple amino
acid substitution model with selection and reversible mutation.}
\end{figure}

For a location where $S=0$, $\pi_{1}^{*}=\beta/(\alpha+\beta)$ and
$\pi_{2}^{*}=\alpha/(\alpha+\beta)$ are the equilibrium amino acid
frequencies, which are solely determined by the mutation pattern.
The expected rate at equilibrium for a neutral location is given by
\begin{equation}
k_{0}=\pi_{1}^{*}\alpha+\pi_{2}^{*}\beta=\frac{2\alpha\beta}{\alpha+\beta}.\label{eq:k_0}
\end{equation}
We can now define the site specific $\omega$ as
\begin{eqnarray}
\omega & = & \frac{k_{s}}{k_{0}}=\frac{\pi_{1}\alpha h(-S)+\pi_{2}\beta h(S)}{\pi_{1}^{*}\alpha+\pi_{2}^{*}\beta}\nonumber \\
 & = & \frac{(1+\frac{\alpha}{\beta})h(S)h(-S)}{h(S)+\frac{\alpha}{\beta}h(-S)}.\label{eq:2x2-omega}
\end{eqnarray}
This equation (\ref{eq:2x2-omega}) describes a typical site under
purifying selection. In most cases $\omega<1$ unless the mutation
rate away from the optimal amino acid is larger than the backward
rate (i.e. $\alpha/\beta>1$). In this case $\omega>1$ might be observed
(figure~\ref{fig:ab-omega}), and the location might spend a longer
fraction of time at $c_{2}$ (\emph{i.e.} $\pi_{2}>\pi_{1}$) despite
this being the suboptimal amino acid. This later scenario seems unlikely
to occur with real data since $\alpha$ and $\beta$ are usually within
the same order of magnitude (\emph{e.g}.~table~6.1 in~\citealt{Lynch2006}).
Note that describing the site as evolving under either `positive'
or `negative' selection might be inappropriate. This is because at
equilibrium the number of positively selected mutant substitutions
equals the number of negatively selected ones (equation~\ref{eq:balance-eq}).

\subsection*{A time non-homogeneous selection model}

It is clear that if selection is constant throughout time no adaptive
evolution can take place. When selection is constant, locations in
a protein will reach a stationary state where the amino acid frequencies
at each location will depend on a balance between random drift, mutation
and selection. We are interested in modelling the case where the selection
coefficient $s$ in favour of amino acid $c_{1}$ is a function of
time. The simplest case is when there is an environment shift, such
as when an organism colonises a new habitat. In this scenario the
previously suboptimal amino acid $c_{1}$ would become optimal, and
the substitution pattern at the location would change. 

Let's write $s_{a}$ for the selection coefficient acting on $c_{1}$
in environment $a$ before the shift, and $s_{b}$ for the coefficient
after a shift towards environment $b$. The pattern of evolution for
the location under each environment can be described by the appropriate
rate matrices:

\begin{equation}
\mathrm{\mathbf{Q}}_{a}=\left(\begin{array}{cc}
-\alpha h(-S_{a}) & \alpha h(-S_{a})\\
\beta h(S_{a}) & -\beta h(S_{a})
\end{array}\right)\mbox{ and }\;\mathrm{\mathbf{Q}}_{b}=\left(\begin{array}{cc}
-\alpha h(-S_{b}) & \alpha h(-S_{b})\\
\beta h(S_{b}) & -\beta h(S_{b})
\end{array}\right)\label{eq:nhomo-simple2x2}
\end{equation}
where $S_{a}=2Ns_{a}$. 

Let's analyse the simple case where $\gamma\equiv\alpha=\beta$, $s_{a}=-s$
and $s_{b}=s$ ($s>0$). That is, $c_{1}$ is the suboptimal amino
acid in $a$ and becomes preferred after the shift towards $b$. Let's
now assume the location has been evolving in $a$ for a very long
time, so it can be considered stationary. The expected substitution
rate immediately after the environment shift ($t=0$) is $k_{s_{b}}(0)=\pi_{1_{a}}\gamma h(-S)+\pi_{2_{a}}\gamma h(S)$,
where $\pi_{i_{a}}$ is the expected equilibrium frequency of $i$
in $a.$ The expected substitution rate in $b$ at equilibrium (\emph{i.e.}
$t\rightarrow\infty)$ is $k_{s_{b}}(\infty)=\pi_{1_{b}}\gamma h(-S)+\pi_{2_{b}}\gamma h(S)$.
In general, we can find the expected substitution rate at any time
point $t$ after the shift, and this is given by

\begin{eqnarray*}
k_{s_{b}}(t) & = & \pi_{1_{a}}p_{11_{b}}(t)q_{12_{b}}+\pi_{1_{a}}p_{12_{b}}(t)q_{21_{b}}\\
 & + & \pi_{2_{a}}p_{22_{b}}(t)q_{21_{b}}+\pi_{2_{a}}p_{21_{b}}(t)q_{12_{b}}
\end{eqnarray*}
where the $p_{ij_{b}}(t)$ are the transition probabilities, which
can be easily found by solving $\mathrm{\mathbf{P}}_{t_{b}}=e^{t\mathrm{\mathbf{Q}}_{b}}$
(\emph{e.g.} p.~39 in \citealt{Yang2006}). For a neutral mutant,
the expected substitution rate is $k_{0}=\gamma$ (equation~\ref{eq:k_0}),
which is independent of $t$ and of whether the population is in $a$
or $b$. So $\omega$ can now be defined as a function of $t$ by
$\omega(t)=\nicefrac{k_{s_{b}}(t)}{k_{0}}$ . With some algebraic
manipulation it is easy to show that 
\begin{equation}
\omega(t)=\omega(\infty)+\left(\omega(0)-\omega(\infty)\right)e^{-\gamma_{s}t}\label{eq:omega-time}
\end{equation}
where $\gamma_{s}=\gamma(h(S)+h(-S))$. For large $S$ we note that
$\omega(0)\approx S$. 

\begin{figure}[p]
\begin{centering}
\includegraphics[scale=0.8]{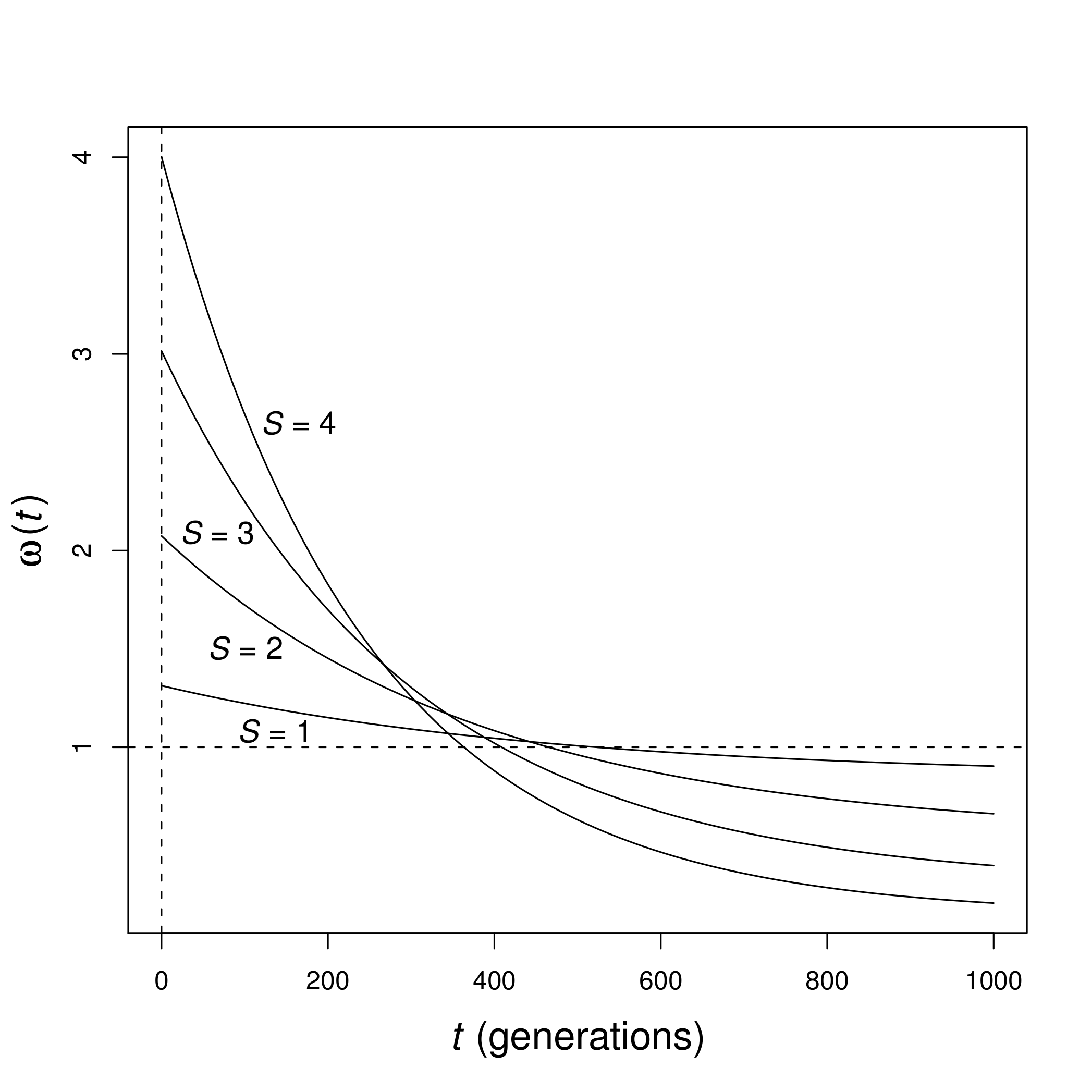}
\par\end{centering}

\caption{\label{fig:omega-time}The expected value of $\omega$ after an environment
shift.\protect \\
The neutral mutation rate is $\gamma=1\times10^{-3}$ per generation.}
\end{figure}

Equation~\ref{eq:omega-time} indicates that immediately after an
environment shift, the substitution rate is accelerated, and $\omega(0)>1$.
As time passes, the system moves towards equilibrium and $\omega(t)$
decays exponentially with rate $\gamma_{s}$ towards $\omega(\infty)<1$
(figure~\ref{fig:omega-time}).

Despite its apparent simplicity, this model has practical applications.
We can imagine a virus that infects and spreads in a certain host
$a$. After a host shift, the intracellular environment in the new
host $b$ might be substantially different, and the viral proteins
would be subjected to novel selective pressures. We shall see later
an actual example with influenza viruses evolving in birds and humans,
and the varying selective pressures involved.

\subsection*{Frequency dependent selection}

The previous models show that for a single location, $\omega$ is
usually less than one unless the mutation bias favours the suboptimal
amino acid (equation~\ref{eq:2x2-omega}) in which case there is
no adaptive evolution involved, or $\omega$ is temporarily larger
than one when there is a shift in the selection pattern acting on
the location (equation~\ref{eq:omega-time}). Several studies, for
example those concerned with the evolution of viral coating proteins,
have shown $\omega$ values for single locations that are consistently
larger than one for large phylogenies (\citealt{Yang2000,Yang+2003}).
Here we seek to describe a population genetics model that can justify
$\omega>1$ at equilibrium. 

We start with a model where the fitness of a mutant allele is a function
of the frequency of the allele in the population. We could imagine,
for example, a virus with a novel mutation in its coating protein
that would allow it to escape the host population immune system. As
the virus would become more common, and host individuals become resistant,
further spreading of the virus would be hampered. Thus, the virus
would lose its selective advantage as its frequency in the population
increases. We can construct a simple model where the fitness ($w$)
of a novel mutant allele, in a haploid organism, decays exponentially
with the frequency ($q$) of the allele: $w(q)=w_{0}e^{-rq}$ where
$w_{0}$ is the initial fitness and $r$ is the rate of decay. The
fitness can be defined in terms of the selection coefficient as $w=1+s$,
so the previous equation can be written as 
\begin{equation}
s(q)=(1+s_{0})e^{-rq}-1,\label{eq:s(p)_exp(-rp)}
\end{equation}
where $s_{0}(>0)$ is the selective advantage of the yet non-existent
mutant (\emph{i.e}. when $q=0$). We are interested in the special
case when $s(1)=0$ (i.e. the fitness of the fixed allele is indistinguishable
from that of the previous wild type), and this is equivalent to setting
$r=\log(1+s_{0})$. Under this condition, equation~\ref{eq:s(p)_exp(-rp)}
becomes
\begin{equation}
s(q)=(1+s_{0})^{1-q}-1.\label{eq:s(p)}
\end{equation}

Let's write $q_{t}$ for the frequency of the allele in the current
generation $t$. In a finite population, $q_{t+1}$ is a binomial
random variable with probability 
\begin{equation}
p(q_{t+1}=i/N)=\left(\begin{array}{c}
N\\
i
\end{array}\right)\rho_{t}^{i}(1-\rho_{t})^{N-i},\label{eq:binom_q_t+1}
\end{equation}
where $N$ is the population number, $\rho_{t}=q_{t}w(q_{t})/\bar{w}(q_{t})$
and $\bar{w}(q_{t})=1+q_{t}s(q_{t})$ is the average fitness (p.~406
in \citealt{Crow+Kimura1970}). Thus, equation~\ref{eq:binom_q_t+1}
can be used to construct the stochastic matrix $\mathrm{\mathbf{P}}=\{p_{ij}\}$
where $p_{ij}$ is the probability that $q_{t}=i/N$ will become $q_{t+1}=j/N$
in one generation ($i,j\in\{0,1,2,\ldots,N\}$). For sufficiently
large $t$, repeated matrix multiplication ($\mathrm{\mathbf{P}}_{t}=\{p_{ij}(t)\}=\Pi^{t}\mathrm{\mathbf{P}}$)
can be used to obtain a numerical approximation to the ultimate fixation
probability, $g(S_{0})\equiv p_{1,N}(t\rightarrow\infty)$, of a novel
mutant with initial frequency $q_{1}=1/N$ and initial selective advantage
$s(1/N)$. It is assumed that $N$ is constant throughout generations,
which might not be a realistic assumption for the case of a virus.

\begin{figure}[p]
\begin{centering}
\includegraphics[scale=0.8]{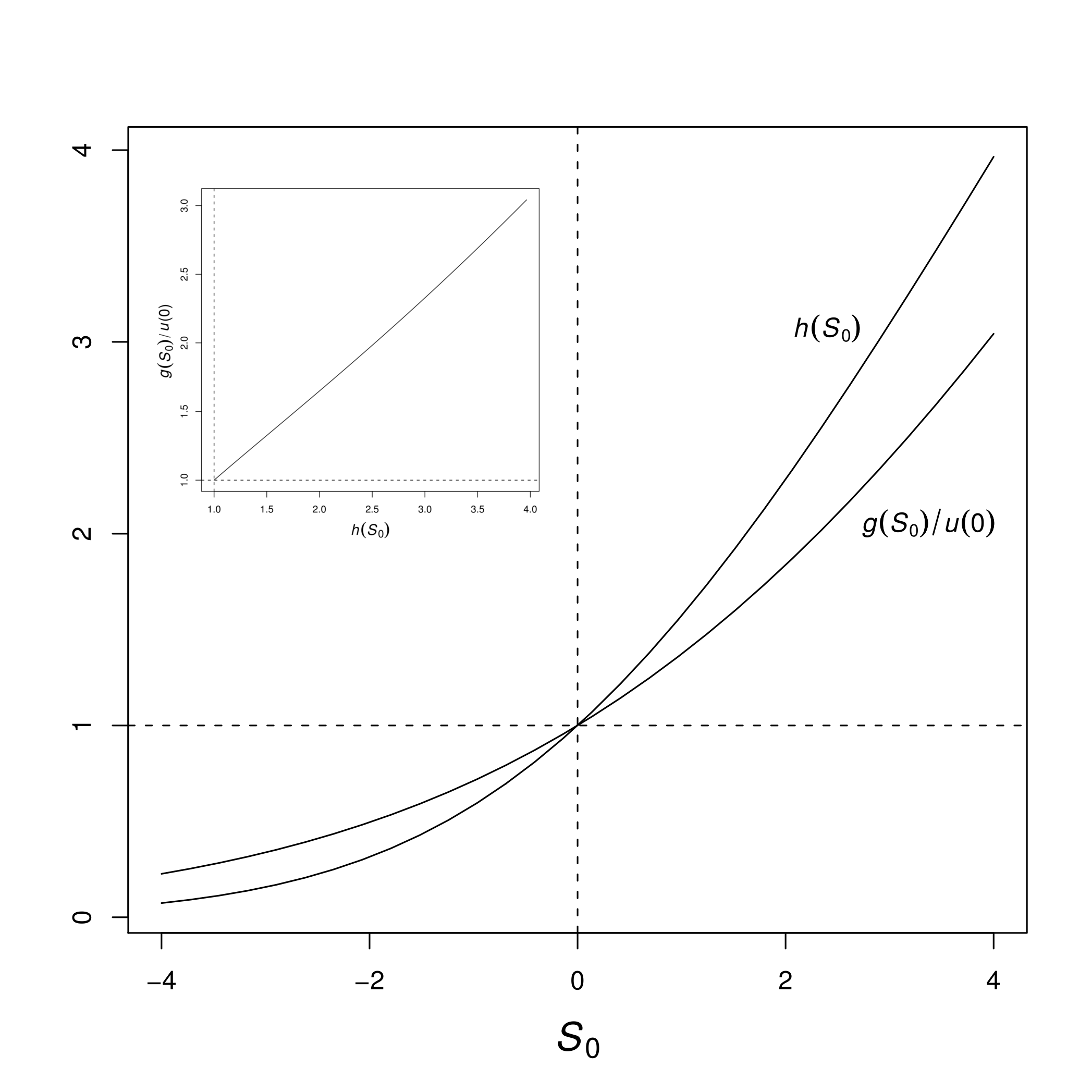}
\par\end{centering}

\caption{\label{fig:h(Z)} The relative fixation probabilities of a novel mutant
under frequency dependent and constant selection.\protect \\
Inset: $g(S_{0})/u(0)$ is approximately a linear function of $h(S_{0})$
for small positive $s_{0}$.}

\end{figure}

Figure~\ref{fig:h(Z)} shows the relative fixation probabilities
$g(S_{0})/u(0)$ and $h(S_{0})\equiv u(S_{0})/u(0)$ as functions
of $S_{0}$ when both are computed numerically (assuming $N=100$).
As should be expected, for $S_{0}>0$ we have that $h(S_{0})>g(S_{0})/u(0)$.
For small positive $s_{0}$, we find numerically that $g(S_{0})/u(0)\approx0.261+0.697h(S_{0})$
(figure~\ref{fig:h(Z)}, inset). We can define numbers $\zeta\equiv\zeta(s_{0})$
and $Z=2N\zeta$ so that 
\[
h(Z)=Z/(1-e^{-Z})\approx g(S_{0})/u(0).
\]
This is, an allele with constant selection coefficient $\zeta$ would
have the same ultimate fixation probability as another with frequency
dependent selection with parameter $s_{0}$. Using the linear relationship
between $h(S_{0})$ and $g(S_{0})/u(0)$ we get that, roughly, $Z\approx0.261+0.697S_{0}$.
I leave the exact computations of $g(S_{0})$ and $Z$ to skillful
mathematicians. The message here is that, when the selective advantage
decays as a function of the allele frequency, the probability of fixation
of the allele is a function of some value $Z$ such that $0<Z<S_{0}$.
This is all we need to construct a substitution model under frequency
dependent selection.

Let's consider again two amino acids $c_{1}$ and $c_{2}$ at a particular
protein location, with mutation rates $\alpha$ and $\beta$. Initially,
the population is fixed for $c_{2}$. The selective advantage of a
novel $c_{1}$ mutant is $s(q)$ where $q$ is the frequency of $c_{1}$
and $s$ is given by equation~\ref{eq:s(p)}. When $c_{1}$ becomes
fixed, its selective advantage is 0 and remains at 0 until it becomes
lost. Returning to our virus example, $c_{1}$ is advantageous when
rare, but once it becomes fixed, the host population has become immunologically
resistant, and will remain so even if the frequency of $c_{1}$ decreases
when a new mutant $c_{2}$ appears. When $c_{2}$ re-appears, the
host population has become again sensitive to $c_{2}$ (\emph{i.e}.
the host population loses its `immune memory' to a virus once this
disappears), and $c_{2}$ has now selective advantage $s(1-q$). The
substitution rates at the location are thus given by

\begin{equation}
\mathrm{\mathbf{Q}}=\left(\begin{array}{cc}
-\alpha h(Z) & \alpha h(Z)\\
\beta h(Z) & -\beta h(Z)
\end{array}\right).\label{eq:Q-freqdepsel}
\end{equation}

The stationary frequencies are $\pi_{1}=\beta h(Z)/(\alpha h(Z)+\beta h(Z))=\beta/(\alpha+\beta)$
and $\pi_{2}=\alpha/(\alpha+\beta)$, so they are independent of the
selective pressure acting on the location. The expected substitution
rate at equilibrium is $k_{s}=\pi_{1}\alpha h(Z)+\pi_{2}\beta h(Z)$.
For a neutral mutant $k_{0}=\pi_{1}^{*}\alpha+\pi_{2}^{*}\beta$,
but in this model $\pi_{i}=\pi_{i}^{*}$ so $\omega$ is simply
\begin{equation}
\omega=\frac{k_{s}}{k_{0}}=h(Z),\label{eq:omega-freqdepsel}
\end{equation}
which is similar to equation~\ref{eq:simple-omega}. 

Setting $\nu=\alpha+\beta$ it can be shown that 

\[
\mathrm{\mathbf{Q}}=\left(\begin{array}{cc}
-\nu\omega\pi_{2} & \nu\omega\pi_{2}\\
\nu\omega\pi_{1} & -\nu\omega\pi_{1}
\end{array}\right),
\]
which is equivalent to the simplest $2\times2$ codon substitution
model when there are no synonymous codons. So in this model we have
that, at equilibrium, $\omega>1$ and $\omega$ is independent of
the mutation pattern.

\section*{Numerical examples with influenza virus data}

\subsection*{Time non-homogeneous selection}

We analyse a set of 401 influenza A polymerase subunit (PB2) sequences
from avian and human isolates. The PB2 alignment and the tree topology
are described by \citet{Tamuri+2009}. The PB2 human isolates are
monophyletic, and they are thought to be the product of a shift from
an avian (the natural reservoir) to a mammalian host dating back to
around 1882\textendash{}1913, before the Spanish flu pandemic of 1918
(\citealt{dosReis+2009}). The PB2 gene codes for a subunit of the
polymerase complex, which is composed of three different subunits
(PB2, PB1 and PA). The polymerase proteins seem to be involved in
host adaptation, and there is evidence of several amino acid substitutions
after the host shift (\citealt{Taubenberger+2005}). 

In this example we focus on amino acid location 627 of the PB2 protein.
Glutamate (E) and lysine (K) are observed in avian isolates, with
glutamate being the predominant amino acid. Lysine is exclusively
observed in the monophyletic human lineage analysed. There is experimental
evidence that lysine at location 627 is related to adaptation to the
mammalian host (\citealt{Subbarao+1993}). Glutamate is encoded by
GAA and GAG codons while lysine is encoded by AAA and AAG codons.
Hence G$\rightleftarrows$A nucleotide transitions are responsible
for E$\rightleftarrows$K mutations. We apply Yang and Nielsen's \citeyearpar{Yang+Nielsen2008}
FMutSel model of codon substitution to the data above to estimate
the background frequencies of G and A nucleotides in the PB2 gene.
These are $\hat{\pi}_{A}^{*}=0.400$ and $\hat{\pi}_{G}^{*}=0.196$
under a HKY85 model of nucleotide substitution (\citealt{Hasegawa+1985}).
The estimated transition transversion rate parameter is $\hat{\kappa}=8.00$.
Thus we estimate the (non-scaled) neutral mutation rates E$\rightarrow$K
as $\hat{\alpha}=\hat{\kappa}\hat{\pi}_{A}^{*}$ and K$\rightarrow$E
as $\hat{\beta}=\hat{\kappa}\hat{\pi}_{G}^{*}$.

We can now construct three substitution models. In the first model
there is a single substitution matrix describing the evolution of
location 627 throughout the tree. The matrix is given by equation~\ref{eq:Q-simple2x2}
and there is no selection ($S=0$). This is the neutral model (M0).
In the next model the substitution matrix is also given by equation~\ref{eq:Q-simple2x2}
but $S(\neq0)$ is now the selective advantage of glutamate. This
is the simple selection model (M1). The final model considers two
substitution matrices, each describing the evolution of the location
in either the avian or the human clades. The background mutational
parameters are the same for both hosts but the selective pressures
in each host are different. We write $s_{\mathrm{av}}$ and $s_{\mathrm{hu}}$
for the selective advantage of glutamate in the avian and human hosts
respectively. The substitution matrices have the form shown in equation~\ref{eq:nhomo-simple2x2}.
The branch linking the human and avian clades is the host shift branch,
where the shift in substitution pattern occurred. This is the non-homogeneous
selection model (M2).

Under each model we can calculate the likelihood of site 627 for the
given tree topology using the pruning algorithm (\citealt{Felsenstein1981,Yang2006}).
We use maximum likelihood to estimate the confounded selection coefficients
($S,S_{\mathrm{av}},S_{\mathrm{hu}}$) for models M1 and M2 for the
fixed tree topology. The branch lengths and mutational parameters
($\alpha$ and $\beta$) are considered fixed, and they are scaled
in terms of neutral mutant substitutions as suggested by \citet{Halpern+Bruno1998}.
Since models M0, M1 and M2 are nested, we use the likelihood ratio
test to select the best model. The estimated parameters for the three
models are shown in table~\ref{tab:flu-simple2x2}. Model M2 is by
far the best model, so we conclude that there are different selective
pressures acting at the location in each host. Note that because $S$
is a confounded parameter of the population number ($N$) and the
selection coefficient ($s$), the differences observed between $S_{\mathrm{av}}$
and $S_{\mathrm{hu}}$ could be due to differences in $N$ as well
as in $s$. However, because there is a change in the \emph{sign}
of $S_{\mathrm{hu}}$, we conclude that glutamate is indeed ill-favoured
in the human host.

\subsection*{A brief comparison of the $\bm{\pi}_{a}\neq\bm{\pi}_{b}$ vs. $\omega>1$
criteria to detect adaptive evolution}

\citet{Tamuri+2009} performed an analysis of selective constraints
in human and avian isolates of influenza. They used a sitewise non-homogeneous
model to test whether amino acid composition in each location of the
influenza proteome would differ between avian and human isolates.
Their model is equivalent to equation~\ref{eq:nhomo-simple2x2},
but the model was parametrised in terms of equilibrium frequencies
and exchangeabilities, as in equation~\ref{eq:Q-nu-pi}, and the
number of amino acids observed per location was usually larger than
two. To account for multiple testing, they used a false discovery
rate (FDR) approach to correct for false positives. For the PB2 gene
they identified 13 locations (FDR$=5\%$, 22 locations for FDR$=20\%$)
where $\bm{\pi}_{\mathrm{hu}}\neq\bm{\pi}_{\mathrm{av}}$ ($\bm{\pi}_{a}=\{\pi_{i_{a}}\}$).
This approach is equivalent to testing if $s_{\mathrm{hu}}\neq s_{\mathrm{av}}$.

The objective now is to compare Tamuri's et al.~\citeyearpar{Tamuri+2009}
$\bm{\pi}_{a}\neq\bm{\pi}_{b}$ criteria to detect whether a site
has undergone adaptive changes, with the classical criteria that $\omega>1$
for the PB2 data. First, a classical codon model where $\omega$ is
constant throughout sites and along the tree was fitted (\citealt{Yang1998}).
A second model, where $\omega$ is allowed to vary between the two
host lineages was also assessed. Allowing $\omega$ to vary between
viruses evolving in human and avian hosts significantly improved the
model fit ($2\Delta\ln\ell=127,\, p\ll0.001$). There is a conspicuous
increase in amino acid substitution rates in human virus isolates
compared to avian isolates ($\hat{\omega}_{\mathrm{av}}=0.0493$ and
$\hat{\omega}_{\mathrm{hu}}=0.1036$). In the traditional sense, no
adaptive evolution has been detected because $\hat{\omega}_{\mathrm{hu}}<1$.
Intuitively, the increase in substitution rates in human viruses would
agree with the notion that these viruses are undergoing adaptive evolution
in the novel host.

Next, we seek to identify locations in the PB2 protein where $\omega>1$.
We use Yang's et al.~\citeyearpar{Yang+2000} M7 and M8 models to
test for $\omega$ variation among codons. In the M7 model, $\omega$
varies following a beta distribution, thus $\omega$ is bounded between
0 and 1. In the M8 model, $\omega$ also varies according to a beta
distribution but there is an extra category of sites with $\omega>1$.
Models M7 and M8 are nested, and M8 has two additional parameters.
The likelihood ratio test can be used to decide whether there are
sites under adaptive evolution, and an empirical Bayes approach can
then be used to identify those sites where $\omega>1$. Fitting M7
and M8 to the human PB2 subtree give essentially the same likelihood,
so the inclusion of a class of sites with $\omega>1$ is not justified.
Under M7, $\hat{\omega}$ varies between 0.0270 and 0.4410 among locations
with a mean value of 0.1076. When M7 and M8 are fitted to all eight
gene segments from human viruses (not shown), the inclusion of positively
selected sites ($\omega>1$) is only justified for the two surface
proteins (HA and NA) and for the nucleoprotein (NP). 

The M8 model averages $\omega$ over a phylogeny, so it might be conservative
when the adaptive event has occurred at a particular branch. It is
possible that most of the adaptive changes for the PB2 protein occurred
along the host shift branch. We apply the branch-sites test of \citet{Zhang+2005}
to the PB2 protein, with the host shift branch linking human and avian
viruses as the `hot' or `foreground' branch, and the rest of the tree
comprising the `brackground' branches. A model allowing a class of
sites with $\omega>1$ in the foreground branch is compared with a
model where $\omega$ is capped at one (\citealt{Zhang+2005}). The
inclusion of a class of sites with $\omega>1$ is not justified for
this branch, as it has the same likelihood as the alternative neutral
model.

\section*{Discussion}

\subsection*{The meaning of $\omega$}

\citet{Sawyer+Hartl1992} provided a theoretical derivation of the
expected fixation rate of synonymous and non-synonymous substitutions
(table~1 in their paper). Although they did not explicitly define
$\omega$, this is perhaps the first work where a formal theoretical
interpretation for this parameter can be found. \citet{Nielsen+Yang2003}
provided an explicit link between a population genetics interpretation
of $\omega$ and Markov models of codon substitution. Using various
models of $\omega$ variation among sites (\citealt{Yang+2000}) they
used equation~\ref{eq:simple-omega} to map the distribution of selection
coefficients from the distribution of $\omega$ ratios in real sequence
data. \citet{Nielsen+Yang2003} assumed a finite sites model, but
the only way to reconcile equation~\ref{eq:simple-omega} with this
assumption is if the fitnesses at a location are reassigned every
time a novel mutant appears, so that a novel mutant has the same selective
advantage (or disadvantage) of the previously fixed allele at the
location. In the case of positively selected mutants, this assumption
is similar to the frequency dependent selection model described here
(equations~\ref{eq:Q-freqdepsel} and~\ref{eq:omega-freqdepsel}).
For negatively selected mutants, this assumption seems unreasonable.
Defining $\omega$ under a purifying selection model (equation~\ref{eq:2x2-omega})
would seem more appropriate. \citet{Kryazhimskiy+Plotkin2008} refer
to the reassignment of fitnesses as the `continual selection' model.
In my opinion, \citet{Nielsen+Yang2003} `reassignment of fitnesses'
seem to be an \emph{ad hoc} justification to use equation~\ref{eq:simple-omega}
to estimate the distribution of selection coefficients, rather than
a prior assumption of their model.

As pointed out by \citet{Kimura1983} {}``One should expect in this
case {[}positive selection{]} that the rate of evolution would depend
strongly on the environment, being high for a species offered a new
ecologic opportunity but low for those kept in a stable environment.''
In the influenza example described here, the novel ecologic opportunity
is given by the introduction of the virus into a new host. In the
novel environment, several locations are suddenly found to be fixed
for suboptimal amino acids and novel mutants have a good chance to
be positively selected. As advantageous mutants appear, spread and
become fixed, the evolutionary rate decays. Eventually, an equilibrium
is reached where most locations become fixed for the optimal state.
Sometimes there might be cyclical changes in the environment, and
locations might undergo recurrent changes in selective pressure. In
the influenza example, the cyclical environment is given by the increasingly
hostile immune system within the host. In this case, a model such
as frequency dependent selection gives a more reasonable interpretation
of the evolutionary rate.

The environment shift model predicts a burst of adaptive evolution
where $\omega$ is initially high ($>1$) and then decays with time
until it reaches a stationary value ($<1$). For the simple $2\times2$
amino acid case, the accelerated substitution rate would be due to
an excess of substitutions occurring at several different locations.
The implications for this are important: although we are likely to
detect an increase in $\omega$ for a clade of interest (such as in
the influenza PB2 example), we might be unable to detect the individual
locations responsible since mostly, all that is observed are single,
one-off substitutions. To detect sites where $\omega>1$, recurrent
substitutions along a branch are necessary. \citet{Tamuri+2009} identified
a total of 172 locations throughout the influenza proteome that show
evidence of adaptive evolution in the human and avian hosts. It is
clear that one-off adaptive substitutions are a widespread phenomenon
in influenza. We can imagine the genome of this virus as composed
of three classes of sites. The first class represents those locations
under strong structural constraints at the protein level, locations
where the optimal amino acid is independent of the host environment.
These locations are described by the constant selection model of equation~\ref{eq:Q-simple2x2}
and represent the largest fraction of the genome. The second class
of locations are those involved in host adaptation, where the substitution
pattern depends on the intracellular environment provided by the particular
host. These locations evolve according to the non-homogeneous selection
model of equation~\ref{eq:nhomo-simple2x2}. Finally, the third class
represents those sites under recurrent changes in selective pressure
and they evolve according to the frequency selection model of equation~\ref{eq:Q-freqdepsel}.
I expect the first two models to account for most of the evolutionary
pattern observed in proteins in most organisms, the second model accounting
for episodic periods of evolution, such as the colonisation of new
habitats or large scale environmental changes. Recurrent selection
models are a special case that would apply to a fraction of locations
in particular proteins.

\subsection*{Gradual change in selective advantage}

Sometimes it might be reasonable to think that the fitness of an allele
would change gradually, perhaps in a correlated manner with some environmental
variable. \citet{Kimura+Ohta1970} studied the fixation probability
of a mutant allele when the selective advantage decreases as a function
of time as $s(t)\propto e^{-rt}$. They showed that the probability
of fixation can be approximated as $Y/N(1-e^{-Y})$ where $Y=2Ny$
is a parameter with a very similar meaning to that of $Z$ in the
frequency dependent selection case. The model of \citet{Kimura+Ohta1970}
could be used to generalise equation~\ref{eq:nhomo-simple2x2} for
the case of gradual environmental changes. However, if the change
in selective pressure is fast compared to expected time between fixations
($1/\mu$) then model~\ref{eq:nhomo-simple2x2} should provide an
adequate approximation.

\subsection*{Pseudo genes and duplicated genes}

\begin{figure}
\begin{centering}
\includegraphics[scale=0.8]{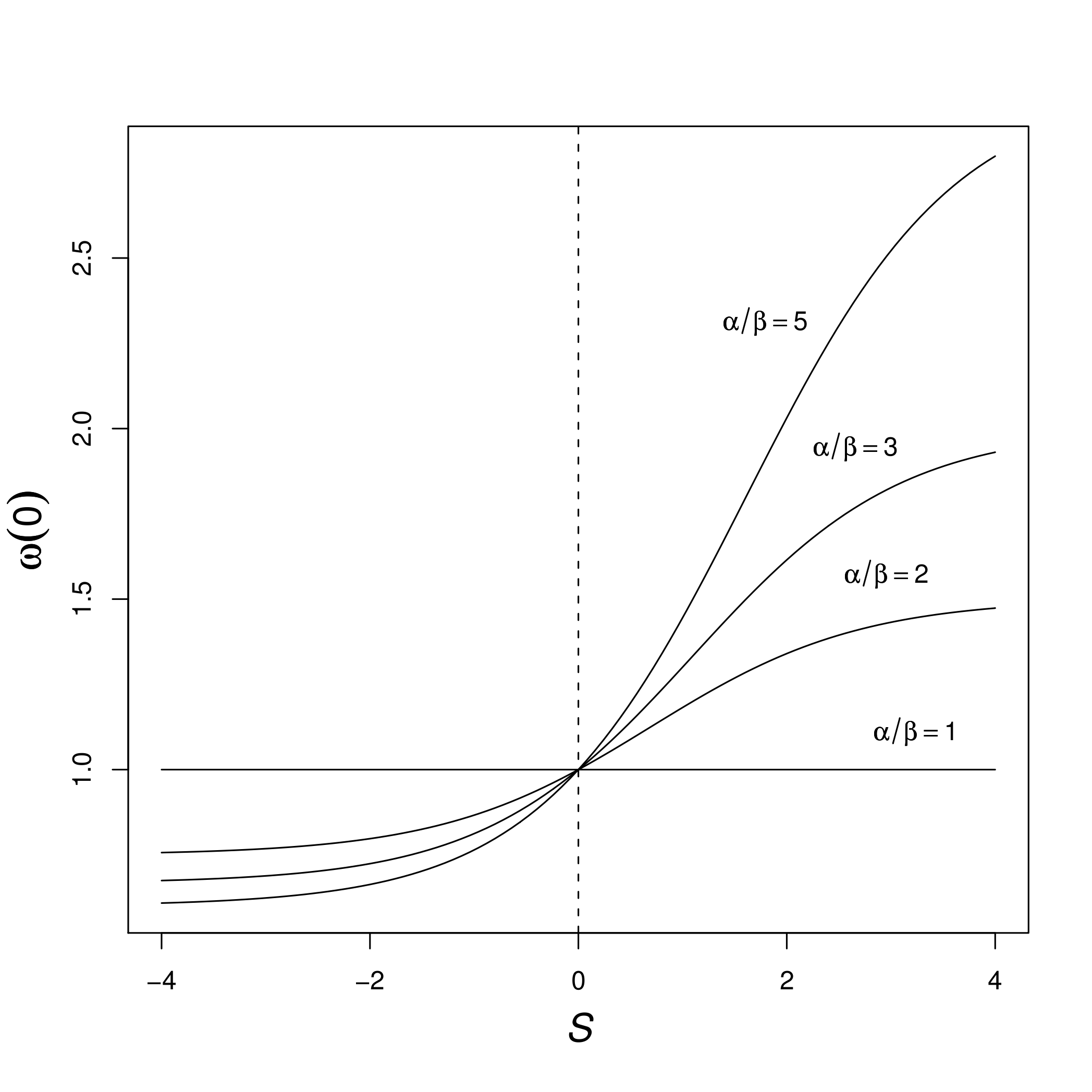}
\par\end{centering}

\caption{\label{fig:w(0)-pseudo}The initial value of $\omega$ for a codon
location in a recently generated pseudo gene.\protect \\
For example, mutational biases A/T$\rightarrow$G/C ($\alpha/\beta$)
of around 4.54 have been observed for some organisms (table~6.1 in~\citealt{Lynch2006}),
considering $S=2.5$, that would mean a codon location with $\omega(0)=2.18$.
On the other hand, if $S=-2.5$, then $\omega(0)=0.65$. A pseudo
gene equally composed of both type of codon locations would evolve
at an average $\omega(0)=0.5\times2.18+0.5\times0.65=1.42$, that
is, substantially faster than the neutral rate at equilibrium.}
\end{figure}

The model described in equation~\ref{eq:nhomo-simple2x2} can also
be used to analyse the substitution rate of a recently generated pseudo
gene. Imagine a gene that becomes duplicated, where the novel copy
lacks a promoter. This situation is equivalent to a sudden environment
shift where $s_{a}>0$ and $s_{b}=0$ (equation~\ref{eq:nhomo-simple2x2}),
then $\mathrm{\mathbf{Q}}_{a}$ represents the substitution pattern
of a location in the functional copy, and $\mathrm{\mathbf{Q}}_{b}$
the pattern of the pseudo gene. With an argument similar to that used
to derive equation~\ref{eq:omega-time}, it can be shown that the
expected value of $\omega$ for the location after the duplication
event ($t=0$) is
\[
\omega(t)=1+(\omega(0)-1)e^{-\nu t},
\]
where $\nu=\alpha+\beta$ and $\omega(0)=(h(S_{a})+h(-S_{a}))(\alpha+\beta)/2\nu_{s}$
($\nu_{s}=\alpha h(-S_{a})+\beta h(S_{a})$). When $\alpha\neq\beta$
and $S>0$, we have that $\omega(0)>1$; when $S<0$ then $\omega(0)<1$;
and when $\alpha=\beta$ then $\omega(0)=1$ whatever the value of
$S$ (figure~\ref{fig:w(0)-pseudo}). As the time after the duplication
approaches infinity, then $\omega(t\rightarrow\infty)\rightarrow1$.
Depending on the degree of mutational bias and the strength of selection
at each location, average $\omega(0)$ for a whole pseudo gene might
exceed one (figure~\ref{fig:w(0)-pseudo}).

If the novel duplicated gene does have a promoter, the organism would
find itself with two functional copies of the same gene. One of the
copies could become dysfunctional if it was to suffer a disruptive
mutation (for example, an enzyme that suffers a critical mutation
at its active site). Since the organism still has a functional copy,
its fitness might be relatively unchanged (for example if the gene
is haplosufficient or double recessive; \citealt{Kondrashov+Koonin2004,Innan2009}).
Then the mutated copy would effectively become a pseudo gene. During
this period the dysfunctional copy would evolve rapidly with $\omega\approx1$,
with some locations reaching $\omega>1$. If due to the serendipitous
nature of evolution this copy was to acquire a novel function, given
its carrier a definite selective advantage, then the substitution
pattern would revert to the case of purifying selection and eventually
$\omega<1$. This observation could explain why $\omega>1$ has been
observed for internal branches of phylogenetic trees purporting gene
duplications. Rather than invoking adaptive evolution, I suspect that
a period of neutral evolution would explain the accelerated rate.
This could be the case, for example, for the primate lysozymes (\citealt{Messier+Stewart1997,Yang1998}).

\subsection*{Models with more than two amino acids}

The $2\times2$ models described here can be easily extended to the
full set of 20 amino acids or 61 codons. For a codon location under
purifying selection, the $61\times61$ substitution matrix could be
defined as

\[
q_{ij}=\begin{cases}
0 & \mbox{if }i\mbox{ and }j\mbox{ differ at more than one position}\\
\mu_{ij} & \mbox{if }i\mbox{ and }j\mbox{ are synonymous}\\
\mu_{ij}h(S_{ij}) & \mbox{if }i\mbox{ and }j\mbox{ are nonsynonymous}
\end{cases},
\]
where $\mu_{ij}$ is the neutral mutation rate from codon $i$ to
$j$, $S_{ij}=2Nf_{j}-2Nf_{i}$ is the confounded selection coefficient
and $f_{j}$ is the fitness of the amino acid encoded by $j$ (\citealt{Yang+Nielsen2008}).
This is very similar to Halpern and Bruno's \citeyearpar{Halpern+Bruno1998}
location specific model to estimate evolutionary distances among coding
sequences. The non-synonymous substitution rate at equilibrium is
$k_{N}=\sum_{i}\sum_{j}\pi_{i}\mu_{ij}h(S_{ij})$ where $i\neq j$
and $aa_{i}\neq aa_{j}$. The expected synonymous rate is $k_{0}=\sum_{i}\sum_{j}\pi_{i}\mu_{ij}$
where $i\neq j$ and $aa_{i}=aa_{j}$. Similarly, when there is no
selection both expected rates are given by $k_{N}^{*}=\sum_{i}\sum_{j}\pi_{i}^{*}\mu_{ij}h(0)$
and $k_{0}^{*}=\sum_{i}\sum_{j}\pi_{i}^{*}\mu_{ij}$. Then $\omega$
can be defined as
\begin{equation}
\omega=\frac{k_{N}}{k_{0}}\times\frac{k_{0}^{*}}{k_{N}^{*}}.\label{eq:final-omega}
\end{equation}
 Terms $k_{N}$, $k_{0}$, $k_{N}^{*}$ and $k_{0}^{*}$ are equivalent
to $\rho_{N}$, $\rho_{S}$, $\rho_{N}^{1}$ and $\rho_{S}^{1}$ in
Yang's~\citeyearpar{Yang2006} notation. Note the similarity of~\ref{eq:final-omega}
with equation~\ref{eq:simple-omega}. In real proteins, structural
and functional constraints restrict the number of amino acids that
can be accommodated into a location, and thus for several amino acids
we would have that $h(S_{ij}\rightarrow-\infty)\rightarrow0$. A consequence
of this is that the numerator in~\ref{eq:final-omega} is reduced
and $\omega$ would tend to be smaller than one. If there is some
form of recurrent selection acting at the location then some of the
$q_{ij}$ terms would be of the form $\mu_{ij}h(Z_{j})$. Because
$Z_{j}>0$ the inclusion of these terms increases the numerator in~\ref{eq:final-omega}
and the value of $\omega$ would increase, possibly over one.

If the time $t$ (in generations) is known, the number of non-synonymous
and synonymous substitutions that have occurred along a lineage are
$tk_{N}$ and $tk_{0}$ respectively. The \emph{scaled} non-synonymous
and synonymous distances are $d_{N}\equiv tk_{N}/k_{N}^{*}$ and $d_{S}\equiv tk_{0}/k_{0}^{*}$.
It is clear that $\omega=d_{N}/d_{S}$, however, because $d_{N}$
and $d_{S}$ are corrected for the relative contributions of non-synonymous
and synonymous mutant substitutions, they should be interpreted with
caution when considered as evolutionary distances.

\subsection*{Future challenges and conclusions}

The models presented here are an approximation to understand $\omega$
at the population genetics level when the substitution process is
considered over long periods of time. Equation~\ref{eq:final-omega}
can be used to construct more realistic evolutionary models that would
include selection as a function of time, as a function of allele frequencies,
or under other models not explored here such as over dominant selection.
\citet{Huelsenbeck+2006} point out that {}``there is {[}currently{]}
little population genetics theory to inform us of the appropriate
probability distribution for among-site variation in the non-synonymous
rate of substitution''. Working out this theory is an important challenge
for both population geneticists and phylogeneticists. If we can understand
the contributions of different modes of selection among locations,
and if we can find an appropriate distribution for the fitnesses of
codons within locations, then a reasonable distribution of $\omega$
values among locations could be worked out. The sitewise approach
exemplified here is computationally expensive and highly parametrised,
and currently is only practical for large phylogenies (such as influenza)
where the pattern of amino acid substitution and the location specific
equilibrium frequencies can be reasonably estimated. Models that use
average, protein wide codon frequencies and that use empirical distributions
for $\omega$ are currently the best approach.

\section*{Note}

This paper was written between January and April 2010.

\section*{Acknowledgements}

Thanks to Lorenz Wernisch, Asif Tamuri, Jens Kleinjung and Richard
Goldstein for helpful comments. This work was supported by the Medical
Research Council, UK.

\bibliographystyle{mbe}
\bibliography{prefs}

\begin{table}[p]
\caption{\label{tab:flu-simple2x2}Adaptive evolution at location 627 of the
PB2 protein of influenza A.}

\begin{centering}
\begin{tabular}{cllccc}
\hline 
Model & $\pi$ & $S$ (95\% CI){*} & np & $\ell$ & $p$-value\tabularnewline
\hline 
\noalign{\vskip\doublerulesep}
M0 & $\hat{\pi}_{E}=0.328$ & $S=0^{**}$ & 0 & $-34.7$ & \tabularnewline
M1 & $\hat{\pi}_{E}=0.668$ & $\hat{S}=+1.4$ ($+0.16,\,+2.9$) & 1 & $-32.2$ & $2.5\times10^{-2}$\tabularnewline
M2 & $\hat{\pi}_{E_{\mathrm{av}}}=0.783$ & $\hat{S}_{\mathrm{av}}=+2.0$ ($+0.30,\,+4.1$) & 2 & $-27.4$ & $1.9\times10^{-3}$\tabularnewline
 & $\hat{\pi}_{E_{\mathrm{hu}}}=0.000$ & $\hat{S}_{\mathrm{hu}}=-940$ ($-\infty,\,-5.5$) &  &  & \tabularnewline[\doublerulesep]
\hline 
 &  &  &  &  & \tabularnewline[\doublerulesep]
\end{tabular}
\par\end{centering}

\begin{centering}
{*} Confidence interval calculated using the LRT statistic (p.~25-26
in~\citealt{Yang2006}). {*}{*} fixed. np: number of parameters estimated
by maximum likelihood.
\par\end{centering}

{\footnotesize NOTE: It is assumed that only two amino acids are allowed
at position 627. In reality, other amino acids could be present at
very low frequencies, and are not observed. A consequence of this
is that the true number of parameters in the models are not known,
and the likelihood ratio tests are thus approximate.}
\end{table}

\end{document}